\documentclass[conference]{IEEEtran}
\pdfoutput=1

\usepackage{graphicx}   
\usepackage{amssymb}
\usepackage{amsmath}
\usepackage{epstopdf}
\usepackage[stretch=16,shrink=16,step=4]{microtype}
\newtheorem{theorem}{Theorem}

\newcommand{\tr}[1]{\ensuremath{{\rm Tr}\lefto(#1\right)}}
\newcommand{\rank}[1]{\ensuremath{{\rm rank}(#1)}}
\newcommand{\dotgeq}{\dot{\geq}}
\newcommand{\dotleq}{\dot{\leq}}
\newcommand{\heff}{\ensuremath{\mathbf{H}_{\rm eff} }}
\newcommand{\prob}[1]{\ensuremath{\mathbb{P}\lefto[#1\right]}}
\newcommand{\capc}[1]{\ensuremath{\frac{1}{2}\log\left(1+#1\right)}}
\newcommand{\innp}[2]{\ensuremath{\langle #1,#2\rangle}}
\newcommand{\lefto}{\mathopen{}\left}
\useRomanappendicesfalse

\makeatletter
\def\@IEEEinterspaceratioM{0.265}
\def\@IEEEinterspaceMINratioM{0.1651}
\def\@IEEEinterspaceMAXratioM{0.38}

\def\@IEEEinterspaceratioB{0.31}
\def\@IEEEinterspaceMINratioB{0.19}
\def\@IEEEinterspaceMAXratioB{0.38}
\@IEEEtunefonts
\makeatother

\IEEEoverridecommandlockouts
\begin{document}
\author{
\begin{tabular}{c}
Cemal Ak\c{c}aba,   Patrick Kuppinger  and  Helmut B\"{o}lcskei 
\end{tabular} \\
Communication Technology Laboratory \\
ETH Zurich, Switzerland \\
Email: \{cakcaba {\textbar} patricku {\textbar} boelcskei\}@nari.ee.ethz.ch
\thanks{This research was supported by Nokia Research Center Helsinki, Finland and by the STREP project No. IST-027310 MEMBRANE within the Sixth Framework Programme of the European Commission.}
}

\title{Distributed Transmit Diversity in Relay Networks} 
\maketitle
\begin{abstract}  
We analyze fading relay networks, where a single-antenna source-destination terminal pair communicates through a set of half-duplex single-antenna relays using a two-hop protocol with linear processing at the relay level. A  family of relaying schemes is presented which achieves the entire optimal diversity-multiplexing (DM) tradeoff curve. As a byproduct of our analysis, it follows that delay diversity and phase-rolling at the relay level are optimal with respect to the entire DM-tradeoff curve, provided the delays and the modulation frequencies, respectively, are chosen appropriately. 
\end{abstract}

\section{Introduction}
Efficiently utilizing the available distributed spatial diversity in wireless networks is a challenging problem.   
In this paper, we consider fading relay networks, where a single-antenna source-destination terminal pair communicates through a set of $K$ half-duplex
single-antenna relays. We assume that there is no direct link between the source and the destination terminals and communication takes place using a two-hop protocol over two time slots. The source terminal and the relays do not have any channel state information (CSI), and the destination terminal knows all channels in the network perfectly.
 
\subsubsection*{Previous work} For setups similar to that described above,  Laneman and Wornell \cite{laneman_st}  propose space-time coded cooperative diversity protocols achieving full spatial diversity gain (i.e., the diversity order equals  the number of relay terminals).  For the setup considered in this paper, Jing and Hassibi \cite{jing06} analyze distributed linear dispersion space-time coding schemes and show that a diversity order equal to the number of relay terminals can be achieved.   In \cite{azarian05},  assuming the presence of a direct link between source and destination, Azarian \emph{et al}.  show that an extension (to the multi-relay case) of a protocol previously introduced in \cite{nabarj} is diversity-multiplexing (DM) tradeoff optimal. 

\subsubsection*{Contributions}  In this paper, we are interested in a class of simple relaying schemes which is based on linear processing at the relay level and hence converts the overall channel between the source and the destination terminal into a time, frequency or time-frequency selective single-input single-output (SISO) channel. This is attractive from an implementation point-of-view, as it allows to realize distributed spatial diversity through the application of standard forward error correction coding over the resulting selective-fading SISO channel.  The class of relaying schemes analyzed in this paper encompasses phase rolling \cite{hammerstroem04, kuppinger06} and cyclic delay diversity \cite{slimane06} at the relay level.  In \cite{slimane06}, it is concluded, through simulations, that a $K$-relay cyclic delay diversity system can achieve a diversity gain of $K$. In \cite{hammerstroem04}, it is demonstrated that phase-rolling at the relay level can achieve second-order diversity. The contributions in this paper can be summarized as follows:
\begin{itemize}
\item{We introduce a broad family of relay transmit diversity schemes based on linear processing at the relay level.}
\
\item{While the (numerical) results in \cite{hammerstroem04,slimane06} are for the case of fixed rate (i.e., the rate does not scale with SNR), we provide a sufficient condition on the proposed class of relay transmit diversity schemes to be optimal with respect to (w.r.t.) the entire DM-tradeoff curve as defined in \cite{zheng_tradeoff}. The tools used to prove DM-tradeoff optimality are a method for computing the optimal DM-tradeoff curve in selective-fading channels, introduced in \cite{coronel06}, and a set of techniques described in \cite{azarian05}.}
\end{itemize}\emph{Notation:}
The superscripts $^{T,H}$ and $^{*}$ stand for transpose, conjugate transpose, and conjugation, respectively.
$x_{i}$ represents the $i$th element of the column vector $\mathbf{x}$, and $\left[\mathbf{X}\right]_{i,j}$ stands for the 
element in the $i$th row and $j$th column of the matrix $\mathbf{X}$. 
$\mathbf{X}\circ \mathbf{Y}$  denotes the Hadamard product of the matrices $\mathbf{X}$ and $\mathbf{Y}$.  $\rank{\mathbf{X}}$ stands for the rank of $\mathbf{X}$.
$\tr{\mathbf{X}}$ and $\| \mathbf{X}\|_{F}$  denote the trace and the Frobenius norm of $\mathbf{X}$, respectively. 
$\mathbf{I}_{N}$ is the $N \times N$ identity matrix. $\mathbf{0}$ denotes the all zeros matrix of appropriate size. We say that the square matrices $\mathbf{X}$ and $\mathbf{Y}$ are orthogonal to each other if $\innp{\mathbf{X}}{\mathbf{Y}} = \tr{\mathbf{X}\mathbf{Y}^{H}}=0$.  All logarithms are to the base $2$. ${\rm diag}(a_{1}, a_{2},\ldots, a_{N})$ denotes the $N \times N$ diagonal matrix with $a_{i}$ on diagonal entry $i$.  The $N \times N$ discrete Fourier transform (DFT) matrix $\mathbf{F}$ is defined as $\left[\mathbf{F}\right]_{ln}= \frac{1}{\sqrt{N}}e^{-j\frac{2\pi}{N}(l-1)(n-1)}$.
$X\sim \mathcal{CN}(0,\sigma^2)$ stands for a circularly symmetric complex Gaussian random variable (RV) with variance $\sigma^{2}$.   Let the positive RV $X$ be parametrized by $\rho>0$. The exponential order of $X$ in $\rho$ is defined as $v = - \frac{\log X}{\log \rho} $. $f(\rho) \doteq g(\rho)$ denotes exponential equality, in $\rho$, of the functions $f(\cdot)$ and $g(\cdot)$, i.e.,  
\[\lim_{\rho\rightarrow\infty} \frac{\log f(\rho)}{\log \rho}  = \lim_{\rho\rightarrow\infty} \frac{\log g(\rho)}{\log \rho} . \]
The symbols $\dotgeq$, $\dotleq$, $\dot{>}$ and $\dot{<}$ are defined analogously.

\section{System Model}
\label{Preliminaries}
\subsubsection*{Preliminaries}
We consider a wireless network with $K+2$ single-antenna terminals, where a source terminal $\mathcal{S}$ communicates with a destination terminal $\mathcal{D}$ through a set of $K$ half-duplex relay terminals $\mathcal{R}_{i}$ $(i=1,2,\ldots,K)$. 
For the sake of simplicity, we assume that there is no direct link between $\mathcal{S}$ and $\mathcal{D}$.   The channels\footnote{$\mathcal{A}\rightarrow\mathcal{B}$ denotes the link between terminals $\mathcal{A}$ and $\mathcal{B}$.} $\mathcal{S}\rightarrow\mathcal{R}_{i}$, with fading coefficient $f_{i}$, and $\mathcal{R}_{i}\rightarrow\mathcal{D}$, with fading coefficient $h_{i}$,  $(i=1,2,\ldots, K)$, are i.i.d. $\mathcal{CN}(0,1)$ and remain constant over the time-scale of interest. We define the column vectors $\mathbf{f } = [f_{1} \ f_{2} \ \cdots\ f_{K}]^{T}$ and $\mathbf{h}=[h_{1} \ h_{2} \ \cdots  \ h_{K}]^{T}$.

Communication takes place over two time slots. In the first time slot, $\mathcal{S}$ transmits $N$ symbols consecutively. The relay terminals process the received length-$N$ sequence using a linear transformation as described in the signal model below and transmit the result during the second time slot to $\mathcal{D}$, while $\mathcal{S}$ remains silent. We assume that $\mathcal{S}$ and the relay terminals do not have CSI, whereas $\mathcal{D}$ knows $f_{i},h_{i}$ $(i=1,2,\ldots, K)$ perfectly. For simplicity, we assume perfect synchronization of the entire network and ignore the impact of shadowing and pathloss. Throughout the paper, we assume that $N\geq K$.

\subsubsection*{Signal model} The vectors $\mathbf{x}$, $\mathbf{r}_{i}$, $\mathbf{y} \in \mathbb{C}^{N}$ represent the transmitted signal, received signal at $\mathcal{R}_{i}$, and received signal at $\mathcal{D}$, respectively. The vector $\mathbf{r}_{i}$ is  given by

\begin{equation}
\mathbf{r}_{i} = \sqrt{\rho}  f_{i} \mathbf{x} + \mathbf{w}_{i}  {\rm , } \ \  i=1,2, \ldots ,K
\end{equation} 
 where $\rho$ denotes the average signal-to-noise ratio (SNR) (for all links)  and  $\mathbf{w}_{i}$ is the $N$-dimensional noise vector at $\mathcal{R}_{i}$, with i.i.d. $\mathcal{CN}(0,1)$ entries. The $\mathbf{w}_{i}$ are independent across $i$ as well.  The transmitted signal $\mathbf{x}$ obeys the constraint $\mathbb{E}\{\mathbf{x}^{H}\mathbf{x}\}= N$.

The relay terminal $\mathcal{R}_{i}$ applies a linear transformation according to $\mathbf{G}_{i}\mathbf{r}_{i}$, where the $N\times N$ matrix $\mathbf{G}_{i}$ satisfies $\mathbf{G}_{i}\mathbf{G}_{i}^{H}= \frac{1}{N}\mathbf{I}_{N}$, scales the result and transmits the signal $ \sqrt{\frac{\rho}{1+\rho}}\mathbf{G}_{i}\mathbf{r}_{i}$.  This ensures that the per-relay transmit power (per dimension) is given by $\rho$. We emphasize that enforcing a per-relay transmit power of $\rho/K$, which leads to a total transmit power (across relays)  of $\rho$,   does not change the main statements and conclusions in the remainder of the paper.   The overall input-output relation reads
\begin{equation}
\label{inputoutputA}
\mathbf{y} = \sum_{i=1}^{K}\frac{\rho}{\sqrt{\rho+1}} \, h_{i}f_{i}\mathbf{G}_{i}\mathbf{x}+ \tilde{\mathbf{z}}
\end{equation}
where the effective noise term $\tilde{\mathbf{z}}$ (when conditioned on $\mathbf{h}$) is circularly symmetric complex Gaussian distributed with $\mathbb{E}\{\tilde{\mathbf{z}}| \mathbf{h}\} = \mathbf{0}$ and 
 $\mathbb{E}\{\tilde{\mathbf{z}}\tilde{\mathbf{z}}^{H}| \mathbf{h}\} = N_{o}^{'}\mathbf{I}_{N}$ where $N_{o}^{'} = \left( 1+{\frac{\rho}{\rho+1}} \|\mathbf{h}\|^{2} \right)$.    
  
Since we will be interested in the mutual information (MI) between $\mathbf{y}$ and $\mathbf{x}$ under the assumption that $\mathcal{D}$ knows all the channels in the network perfectly, we can divide (\ref{inputoutputA}) by $\sqrt{N_{o}^{'}}$ to obtain the effective input-output relation

\begin{equation}
\label{inputoutputD}
\mathbf{y} = \frac{\rho}{\sqrt{1+\rho(1+\|\mathbf{h}\|^2)}}\sum_{i=1}^{K} h_{i}f_{i}\mathbf{G}_{i}\mathbf{x}  +\mathbf{z} 
\end{equation}
where  $\mathbf{z}$ (when conditioned on $\mathbf{h}$) is a circularly symmetric complex Gaussian noise vector with $\mathbb{E}\{{\mathbf{z}}| \mathbf{h}\} = \mathbf{0}$ and 
 $\mathbb{E}\{{\mathbf{z}}{\mathbf{z}}^{H}| \mathbf{h}\} = \mathbf{I}_{N}$.     In the remainder of the paper, we shall be interested in the $\rho\rightarrow\infty$ case where $\frac{\rho}{\sqrt{1+\rho(1+\|\mathbf{h}\|^2)}} \approx \sqrt{\frac{\rho}{1+\|\mathbf{h}\|^2}}$.  With $\heff = \frac{1}{\sqrt{1+||\mathbf{h}||^2}}\sum\limits_{i=1}^{K}h_{i}f_{i}\mathbf{G}_{i}$, we can now rewrite the input-output relation (\ref{inputoutputD}) as 
\begin{equation}
\label{inputoutputD1}
\mathbf{y} = \sqrt{\rho} \, \heff\mathbf{x} +\mathbf{z} {\rm. }
\end{equation}

\section{Achieving The Optimal Diversity-Multiplexing Tradeoff}
Under the assumptions stated in the previous section, it follows that  the maximum MI of the effective channel in (\ref{inputoutputD1}) is achieved by i.i.d. Gaussian codebooks.  
The corresponding MI is given by
\begin{equation}
\label{mutualinfo}
I(\mathbf{y};\mathbf{x}|  \heff) = \frac{1}{2N} \sum_{n=0}^{N-1}\log (1+ \rho\lambda_{n} (\heff \heff^{H}))
\end{equation}
where the factor $1/2$ is due to the half-duplex constraint. 

The DM-tradeoff realized by a family (one at each SNR $\rho$) of codebooks $\mathcal{C}_{r}$  with rate $R=r\log\rho$, where $r \in [0, 1/2]$, is given by the function
\begin{align}
d(r) = -\lim_{\rho\rightarrow\infty} \frac{\log P_{e}(\rho,r)}{\log\rho} \notag
\end{align}
where $P_{e}(\rho,r)$ is the error probability obtained through maximum likelihood (ML) decoding. We say that $\mathcal{C}_{r}$ operates at multiplexing gain $r$.
For a given SNR $\rho$, the codebook $\mathcal{C}_{r}(\rho) \in \mathcal{C}_{r}$ contains $\rho^{2Nr}$ codewords $\mathbf{x}_{i}$. 

Next, we compute the optimal DM-tradeoff curve, as defined in \cite{zheng_tradeoff}, for the effective channel $\heff$ and provide a sufficient condition on the matrices $\mathbf{G}_{i}$ $(i=1,2,\dots,K)$ in conjunction with a family of codebooks $\mathcal{C}_{r}$ $(r\in [0, 1/2])$ to be DM-tradeoff optimal.  Following the framework in \cite{zheng_tradeoff},  we define the probability of outage at multiplexing gain $r$ and SNR $\rho$ as 
\begin{equation}
\label{outageprob}
P_{\mathcal{O}}(\rho,r) = \prob{I(\mathbf{y};\mathbf{x}|\heff) < r\log\rho}  {\rm .}
\end{equation}
Directly analyzing (\ref{outageprob}) is challenging as closed-form expressions for the eigenvalue distribution of $\heff$ do not seem to be available.  However, noting that
\begin{align}
\label{jensenmutualinfo2}
I(\mathbf{y}; \mathbf{x} | \heff) &\leq  I_{J}(\mathbf{y};\mathbf{x} | \heff)  
\end{align}
where
\begin{align}
I_{J}(\mathbf{y};\mathbf{x} | \heff) & =    \frac{1}{2}\log \left(1+ \frac{\rho}{N} \sum_{n=0}^{N-1}\lambda_{n} (\heff \heff^{H})\right) \notag \\
\label{jensenmutualinfo}
& = \frac{1}{2}\log\lefto(1+\frac{\rho}{N}\|\heff\|_{F}^{2}\right)
\end{align} we can resort to a technique developed in \cite{coronel06} to show that the DM-tradeoff corresponding to $I_{J}(\mathbf{y};\mathbf{x} | \heff)$ equals that corresponding to $I(\mathbf{y}; \mathbf{x} | \heff)$. The significance of this result lies in the fact that the quantity $\| \heff \|_{F}^{2}$ lends itself nicely to analytical treatment. 

 \label{Sufficiency Conditions}
In the following, we will need the $N \times K$ code difference matrix defined as
\begin{align}
\mathbf{\Phi}(\Delta \mathbf{x})=\left[\mathbf{G}_{1}\Delta\mathbf{x} \ \  \mathbf{G}_{2}\Delta\mathbf{x} \ \ \cdots \ \ \mathbf{G}_{K}\Delta\mathbf{x} \right]
\end{align}
where $\Delta\mathbf{x} = \tilde{\mathbf{x}}-\hat{\mathbf{x}}$ denotes the code difference vector associated with the codewords $\tilde{\mathbf{x}}, \hat{\mathbf{x}}$. Our main result can now be summarized as follows. 
    
\begin{theorem} 
\label{th-jensen}
For the half-duplex relay channel in (\ref{inputoutputD1}), the optimal DM-tradeoff curve is given by
\begin{equation}
d(r)= K(1-2r), \ r \in [0,1/2].
\end{equation}
Let $\{\mathbf{G}_{1}, \mathbf{G}_{2}, \dots, \mathbf{G}_{K} \}$ be a set of transformation matrices and $\mathcal{C}_{r}$ a family of codebooks such that for any codebook $\mathcal{C}_{r}(\rho) \in\mathcal{C}_{r}$ and any two codewords $\tilde{\mathbf{x}}, \hat{\mathbf{x}} \in \mathcal{C}_{r}(\rho)$ the condition
$\rank{\mathbf{\Phi}(\Delta\mathbf{x})}=K$ holds. Then, the ML decoding error probability satisfies
\begin{align}
P_{e}(\rho,r) \doteq \rho^{-d(r)}.
\end{align} 
\end{theorem}  

 \begin{proof}
See Appendix \ref{ProofofJensen}.
\end{proof}

\emph{Discussion:}
Theorem \ref{th-jensen} shows that the DM-tradeoff properties of the half-duplex relay channel in (\ref{inputoutputD1}) are equal to the ``cooperative upper bound'' (apart from the factor $1/2$ loss, which is due to the half-duplex constraint) corresponding to a system with one transmit and $K$ cooperating receive antennas. Noise forwarding at the relay level and the lack of cooperation, hence, do not impact the DM-tradeoff behavior, provided the matrices $\mathbf{G}_{i}$ and the family of codebooks $\mathcal{C}_{r}$ are chosen according to the conditions in Theorem 1.  Azarian \emph{et al.} \cite{azarian05}, assuming the presence of a direct link between source and destination, show that extending Protocol I in \cite{nabarj} to the multi-relay case by allowing only one relay to transmit in a given time slot yields DM-tradeoff optimality w.r.t. the entire DM-tradeoff curve.  Our results show, however, that DM-tradeoff optimality can be obtained even if all relays transmit in all time slots as long as the full-rank condition in Theorem 1 is satisfied.  Another immediate conclusion that can be drawn from Theorem \ref{th-jensen} is that cyclic delay diversity \cite{slimane06} and phase-rolling \cite{hammerstroem04,kuppinger06} at the relay level are optimal w.r.t. the entire DM-tradeoff curve, provided the delays, the modulation frequencies and the codebooks  are chosen appropriately. This can be seen as follows. We start by noting that the cyclic delay diversity scheme \cite{slimane06} can be cast into our framework by setting $\mathbf{G}_{i}= \frac{1}{\sqrt{N}}\mathbf{P}_{i}$ where $\mathbf{P}_{i}$ denotes the permutation matrix that, when applied to a vector $\mathbf{x}$, cyclically shifts the elements in $\mathbf{x}$ up by $i-1$ positions.With 

\begin{equation}
\label{ppaz}
\innp{\mathbf{P}_{i}}{\mathbf{P}_{j}} =
  \begin{cases}
    N, & {\rm  \ } i = j  \\ 
    0, & {\rm   \ } i \neq j
    \end{cases}     
\end{equation}
the condition $\rank{\mathbf{\Phi}(\Delta \mathbf{x})}=K$ takes a particularly simple form, namely $(\mathbf{F}{\Delta\mathbf{x}})_{k}\neq 0$  for all $k \in \{1, 2, \ldots,N\}$.  To see this note that $\rank{\mathbf{\Phi}(\Delta \mathbf{x})}=\rank{\mathbf{F}\mathbf{\Phi}(\Delta \mathbf{x})}$ and $\mathbf{P}_{i} =\mathbf{F}^{H}\mathbf{\Lambda}_{i}\mathbf{F}$ where
\begin{align}
\mathbf{\Lambda}_{i} = {\rm diag}\lefto(e^{j\theta_{i}[0]}, e^{j\theta_{i}[1]}, \cdots,   e^{j\theta_{i}[N-1]}\right)
\end{align}
with $\theta_{i}[n] = \frac{2\pi n (i-1)}{N}$. Next, we have 
\begin{align}
\label{aspar}
\rank{\mathbf{F}\mathbf{\Phi}(\Delta \mathbf{x})} = \rank{\mathbf{\Sigma}\left[ \mathbf{l}_{1} \ \mathbf{l}_{2} \ \cdots \  \mathbf{l}_{K} \right]}
\end{align}
where $\mathbf{\Sigma}= {\rm diag}\left((\mathbf{F}{\Delta \mathbf{x}})_{1},(\mathbf{F}{\Delta \mathbf{x}})_{2}, \ldots, (\mathbf{F}{\Delta \mathbf{x}})_{N}\right)$ and $[\mathbf{l}_{i}]_{k} \! = e^{j\theta_{i}[k-1]}, k = 1, 2, \ldots,N$,  $i = 1, 2, \ldots K$. As a consequence of (\ref{ppaz}), the columns of the matrix $\left[ \mathbf{l}_{1} \ \mathbf{l}_{2} \ \cdots \  \mathbf{l}_{K} \right]$ are orthogonal and hence $\rank{\mathbf{F}\mathbf{\Phi}(\Delta \mathbf{x})}=K$  if $\mathbf{\Sigma}$ has full rank which is the case if $\left(\mathbf{F}\Delta\mathbf{x}\right)_{k} \neq 0$  for all $k \in \{1, 2, \ldots,N\}$.

In the case of phase-rolling \cite{hammerstroem04, kuppinger06}, we have $ \mathbf{G}_{i} = \frac{1}{\sqrt{N}}\mathbf{\Lambda}_{i}$.  Again, the condition $\rank{\mathbf{\Phi}(\Delta\mathbf{x})}=K$ takes a particularly simple form, namely $(\Delta\mathbf{x})_{k}\neq0$ for all $k \in \left\{1, 2, \ldots,N\right\}$.  The proof of this statement follows by considering $\mathbf{\Phi}(\Delta\mathbf{x})$ directly, putting $\rank{\mathbf{\Phi}(\Delta\mathbf{x})}$ into the form of the right-hand side of (\ref{aspar}) and applying the remaining steps in the argument for the cyclic delay diversity case. While the (numerical) results in \cite{slimane06,hammerstroem04} are for the $r=0$ case, our analysis reveals optimality of cyclic delay diversity and phase-rolling for the entire DM-tradeoff curve, provided the codebooks satisfy the full-rank condition in Theorem 1.  We finally note that cyclic delay diversity and phase-rolling are time-frequency duals of each other in the sense that the linear transformation matrices for the two schemes obey $\mathbf{G}_{i}=\frac{1}{\sqrt{N}}\mathbf{F}\mathbf{P}_{i}\mathbf{F}^{H}$. 
\subsubsection*{Relation to approximately universal codes \cite{tavildar06}}
For the half-duplex relay channel investigated in this paper, a family of codes $\mathcal{C}_{r}$ is DM-tradeoff optimal if 
\begin{align}
\label{auc}
\mu_{\min}(\rho) \  \dot{>} \ \rho^{-2r}
\end{align}
where $\mu_{\min}(\rho)$ is the smallest eigenvalue of $(\mathbf{\Phi}(\Delta\mathbf{x}))^{H}\mathbf{\Phi}(\Delta\mathbf{x})$ over all $\Delta\mathbf{x}=\tilde{\mathbf{x}}-\hat{\mathbf{x}}$ with $\tilde{\mathbf{x}},\hat{\mathbf{x}} \in \mathcal{C}_{r}(\rho)$.  This result follows immediately from (\ref{error-no-jensen}) in the proof of Theorem 1. Based on (\ref{auc}), we can conclude (using the same arguments as in Sec. IV. A in \cite{coronel06}) that any family of codes $\mathcal{C}_{r}$ satisfying (\ref{auc}) will also be approximately universal in the sense of \cite[Th. 3.1]{tavildar06}. 
\subsubsection*{Relation to code design criteria for point-to-point case} 
We conclude our discussion by pointing out that the conditions of Theorem 1 guarantee DM-tradeoff optimality in point-to-point multiple-input single-output systems as well. 

\section{Conclusions}
We introduced a family of linear relay processing schemes achieving the optimal DM-tradeoff curve of half-duplex relay channels. Cyclic delay diversity and phase-rolling were shown to be (DM-tradeoff optimal) special cases. Our analysis can readily be extended to account for the presence of a direct link between the source and the destination terminals. Finally, we note that the DM-tradeoff framework seems to be too crude to quantify potential performance differences between relay transmit diversity schemes with different eigenvalue spread of the Gramian matrix of the $\mathbf{G}_{i}$. 

\appendices
\section{Proof of Theorem 1}
\label{ProofofJensen}
We start by noting that an upper bound on the DM-tradeoff curve can be obtained by applying the broadcast cut-set bound \cite{gastpar05} to the described network 
and evaluating the corresponding DM-tradeoff for i.i.d. Gaussian codebooks. It is shown in \cite{gastpar05, Bol06} that the broadcast cut amounts to a point-to-point link with a single transmit and $K$ (cooperating) receive antennas. Taking into account the factor $1/2$ loss due to the half-duplex nature of the relay terminals, it follows immediately from the results in \cite{zheng_tradeoff} that the DM-tradeoff curve corresponding to the network analyzed in this paper is upper-bounded by
\[ d(r) \leq K(1-2r), \ r \in [0,1/2].\]
In the following, we shall show that this upper bound is achievable, despite the lack of cooperation between the relay terminals, provided that, for every $r \in [0,1/2]$,  $\mathcal{C}_{r}(\rho) \in \mathcal{C}_{r}$ satisfies $\rank{\mathbf{\Phi}(\Delta\mathbf{x})}=K$ for all $\Delta\mathbf{x} = \tilde{\mathbf{x}}-\hat{\mathbf{x}}$ with $\tilde{\mathbf{x}},\hat{\mathbf{x}}\in\mathcal{C}_{r}(\rho)$.  
We start by noting that
\begin{equation}
P_{\mathcal{O}}(\rho,r)\geq \prob{\mathcal{J}} = \prob{I_{J}(\mathbf{y};\mathbf{x}|\heff) < r\log\rho}  \notag
\end{equation}
where 
\begin{equation}  \mathcal{J}= \left\{\heff \lvert   I_{J}(\mathbf{y};\mathbf{x} | \heff) < r\log\rho \right\} \notag \end{equation} is defined as ``Jensen outage'' event.   Since 
\begin{align}
 \frac{\rho}{N}\| \heff \|_{F}^{2} & = \frac{\rho}{N(1+\|\mathbf{h}\|^2)}\tr{\sum\limits_{i=1}^{K}\sum_{j=1}^{K}(h_{i}f_{i})(h_{j}^{*}f_{j}^{*})\mathbf{G}_{i}\mathbf{G}_{j}^{H}} \notag \\
& = \frac{\rho}{1+\| \mathbf{h}\|^{2}} \tilde{\mathbf{h}}^{H}\mathbf{K}\tilde{\mathbf{h}} \notag
 \end{align}
where $\tilde{\mathbf{h}} = \mathbf{h}\circ\mathbf{f}$ and the Gramian 
\begin{equation} 
\label{GramianX1}
\mathbf{K}= 
\frac{1}{N}\lefto[
\begin{array}{ccc}
 \tr{\mathbf{G}_{1}\mathbf{G}_{1}^{H}} & \cdots   & \tr{\mathbf{G}_{K}\mathbf{G}_{1}^{H}}   \\
 \vdots & \vdots & \vdots \\
 \tr{\mathbf{G}_{1}\mathbf{G}_{K}^{H}} & \cdots   & \tr{\mathbf{G}_{K}\mathbf{G}_{K}^{H}}   
\end{array}
\right]\notag
\end{equation}
we have 
\begin{align}
\prob{\mathcal{J}} =  \mathbb{P}\left[\frac{1}{2}\log\left(1+\rho\frac{\tilde{\mathbf{h}}^{H}\mathbf{K}\tilde{\mathbf{h}}}{1+\|\mathbf{h}\|^{2}} \right) < r\log \rho\right]  \notag
 .\end{align}

In what follows, we write $|h_{i}|^2 = \rho^{-u_{i}}$ and $|f_{i}|^2 = \rho^{-v_{i}}$  where $u_{i}$ and $v_{i}$ are RVs; the choice
of this transformation will become clear later.  Further, we define the events
 $\mathcal{A}=\{u_{1}, u_{2},\ldots,u_{K}, v_{1}, v_{2},\ldots, v_{K}| u_{i} \geq 0, v_{i} \geq 0 \ \forall i \in \{1,2,\ldots, K\}\}$ and the complementary event $\bar{\mathcal{A}}$ as the event where at least one $u_{i}$ or $v_{i}$ is negative. 
Using the law of total probability, we can write 
 \begin{align}
 \prob{\mathcal{J}} = \prob{\mathcal{A}}\prob{\mathcal{J}|\mathcal{A}} + \prob{\bar{\mathcal{A}}}\prob{\mathcal{J}|\bar{\mathcal{A}}} \notag
 \end{align}
and bound $\prob{\mathcal{J}}$ according to
 \begin{align}
 \prob{\mathcal{A}}\prob{\mathcal{J}|\mathcal{A}}   \leq & \ \prob{\mathcal{J}}   \leq \prob{\mathcal{A}}\prob{\mathcal{J}|\mathcal{A}} + \prob{\bar{\mathcal{A}}} \notag \\
 \label{PA}\prob{\mathcal{A}}\prob{\mathcal{J}|\mathcal{A}} \dotleq    & \ \prob{\mathcal{J}}\  \dotleq \ \prob{\mathcal{A}}\prob{\mathcal{J}|\mathcal{A}}  \\
 \label{PA2}\prob{\mathcal{J}|\mathcal{A}} \dotleq &  \  \prob{\mathcal{J}} \ \dotleq \ \prob{\mathcal{J}|\mathcal{A}}
 \end{align}
where  (\ref{PA}) follows from the definition of the $u_{i}$ and the $v_{i}$, their independence and by noting that
$\prob{\bar{\mathcal{A}}}$ decays exponentially fast in $\rho$.  The double inequality (\ref{PA2}) results  from $\lim\limits_{\rho\rightarrow\infty}\frac{\log\prob{\mathcal{A}}}{\log\rho} = 0$. We have thus shown that $\prob{\mathcal{J}}\doteq \prob{\mathcal{J}\lvert\mathcal{A}}$. Next, denoting the minimum and maximum eigenvalue of $\mathbf{K}$ as $\lambda_{{\rm min}}$ and $\lambda_{{\rm max}}$, respectively, we get the upper bound
\begin{align}
\label{PA3}&\prob{\mathcal{J}|\mathcal{A}} \  \dotleq \ \prob{\capc{\rho\frac{\lambda_{{\rm min}} }{1+K} \|\tilde{\mathbf{h}}\|^2 }<r\log\rho } \\
\ & = \prob{\capc{\rho\frac{\lambda_{{\rm min}} }{1+K} \sum_{i=1}^{K} |f_{i}|^2|h_{i}|^{2}}<r\log\rho } \notag \\
\ & = \prob{\capc{\frac{\lambda_{\rm min}}{1+K} \sum\limits_{i=1}^{K} \rho^{1-v_{i}-u_{i}}}<r\log\rho } \notag
\end{align}  and the lower bound
\begin{align}
\label{PA4}&\prob{\mathcal{J}|\mathcal{A}} \  \dotgeq   \  \prob{\capc{\rho\lambda_{{\rm max}}  \|\tilde{\mathbf{h}}\|^2}<r\log\rho } \\
\ & = \prob{\capc{\rho\lambda_{{\rm max}} \sum_{i=1}^{K} |f_{i}|^2|h_{i}|^{2}}<r\log\rho } \notag \\
\ & = \prob{\capc{\lambda_{{\rm max}}  \sum\limits_{i=1}^{K} \rho^{1-v_{i}-u_{i}}}<r\log\rho }  \notag
\end{align}
where the key steps (\ref{PA3}) and (\ref{PA4}) follow from the Rayleigh-Ritz theorem \cite{Horn85} and the fact that  
$ 1 \leq 1+ \sum\limits_{i=1}^{K}\rho^{-u_{i}} \leq 1+K$ for  $u_{i}\geq0$ $(i=1,2,\ldots, K)$ and  $\rho>1$.  It can be shown that the full-rank condition on $\mathbf{\Phi}(\Delta\mathbf{x})$ implies $\rank{\mathbf{K}}=K$ and therefore $\lambda_{{\rm min}} > 0$. We next define the following events 
\begin{align}
\mathcal{B} &= \lefto\{ u_{i}, v_{i} \lvert \max_{ i} (1-v_{i}-u_{i}) > 0   \right\} \notag \\ 
\mathcal{U} &= \lefto\{ u_{i}, v_{i}  \left| \capc{\frac{\lambda_{{\rm min}}\rho^{\max_{i}(1-v_{i}-u_{i})}}{1+K}}<r\log\rho \right. \right\} \notag \\ 
\mathcal{L} &= \lefto\{u_{i}, v_{i} \left|  \capc{K\lambda_{{\rm max}}\rho^{\max_{i}(1-v_{i}-u_{i})}}<r\log\rho \right. \right\} \notag 
\end{align}
where the $\max$ is taken over $i=1,2,\ldots, K$ in all three cases. With these definitions, we arrive at
\begin{align}
\label{PCC1} \prob{\mathcal{L}}\dotleq \ &\prob{\mathcal{J}|\mathcal{A}} \dotleq \ \prob{\mathcal{U}} \\
\prob{\mathcal{L}\cap\mathcal{B}} + \prob{\mathcal{L}\cap\bar{\mathcal{B}}}\dotleq \ &\prob{\mathcal{J}|\mathcal{A}} \dotleq \ \prob{\mathcal{U}\cap\mathcal{B}} + \prob{\mathcal{U}\cap\bar{\mathcal{B}}} \notag \\
 \prob{\mathcal{L}\cap\mathcal{B}} \dotleq \ &\prob{\mathcal{J}|\mathcal{A}} \dotleq \ \prob{\mathcal{U}\cap\mathcal{B}} + \prob{\bar{\mathcal{B}}} \notag 
\end{align} 
where (\ref{PCC1}) follows from 
\begin{align}
  1 \leq \frac{\sum_{i=1}^{K} \rho^{1-v_{i}-u_{i}}}{ \rho^{\max_{i } (1-v_{i}-u_{i})} } \ & \leq \ K \notag.
\end{align}
Now, we can expand $ \prob{\mathcal{U}\cap\mathcal{B}}$ as 
 \begin{align}
\label{PXX01} & \prob{\mathcal{U}\cap\mathcal{B}}  =   \ \prob{0<\max_{i}(1-v_{i}-u_{i}) < 2r + \epsilon_{1}} \\
&   =\lefto(\prob{ |f_{1}|^2|h_{1}|^2 \! \! < \frac{1}{{\rho^{1-2r-\epsilon_{1}}}}}\right)^{K} \! \! \! \! \!   -\lefto(\prob{ |f_{1}|^2|h_{1}|^2 \! \!< \frac{1}{{\rho}}}\right)^{K} \notag \\
& \label{PXX22} = \left(1 - {\rm F}_{1}\! \lefto(\!{\sqrt{\rho^{1-2r-\epsilon_{1}}}}\right)\right)^K  -  \lefto(1- {\rm F}_{1}\! \left({\sqrt{\rho}}\right)\right)^{K}
\end{align} 
where $\epsilon_{1} = \log(\frac{1+K}{\lambda_{{\rm min}}})/{\log \rho}$ and ${\rm F}_{1}(x) = \frac{2}{x}{\rm K}_{1}(\frac{2}{x})$ with ${\rm K}_{1}(\cdot)$ denoting the first-order modified Bessel function of the second kind. Further, we have 
\begin{align}
\label{PXX23}\prob{\bar{\mathcal{B}}} = \lefto(1- {\rm F}_{1}\!\left(\!{\sqrt{\rho}}\right)\right)^{K} 
\end{align}
where for (\ref{PXX22}) and (\ref{PXX23}) we used the fact that the CDF of the product of two Rayleigh distributed RVs is given by $1-2x{\rm K}_{1}(2x)$ for $x>0$ \cite{Salo06}. In the ensuing discussion, all statements involving $r$ hold for $r \in [0,1/2]$.

Combining  (\ref{PXX22}) and (\ref{PXX23}), we get  
\begin{align}
\prob{\mathcal{U}\cap\mathcal{B}} + \prob{\bar{\mathcal{B}}} = &  \left(1 - {\rm F}_{1}\lefto({\sqrt{\rho^{1-2r-\epsilon_{1}}}}\right)\right)^K\notag \\ 
\label{PXX1} \doteq & \ \rho^{-K(1-2r)}
 \end{align}
 where the exponential equality in (\ref{PXX1}) is proved using a Taylor series expansion of $1-{\rm F}_{1}(1/x)$ around $x=0$ and invoking asymptotic properties of $\log(x)$ \cite[Eq. (4.1.30-31)]{Abr65}. To complete the proof, we establish that $ \prob{\mathcal{L}\cap\mathcal{B}}$ has the same exponential behavior (in $\rho$) as  $ \prob{\mathcal{U}\cap\mathcal{B}}+\prob{\bar{\mathcal{B}}}$. Using the same arguments as in (\ref{PXX01})-(\ref{PXX1}), it readily follows that
\begin{align}
\prob{\mathcal{L}\cap\mathcal{B}}=  \prob{0<\max_{i}(1-v_{i}-u_{i})<2r-\epsilon_{2}} \notag \\
\	 	= \left(1 - {\rm F}_{1}\!\lefto({\sqrt{\rho^{1-2r+\epsilon_{2}}}}\right)\right)^K  \! \!-\lefto(1- {\rm F}_{1}\!\left({\sqrt{\rho}}\right)\right)^{K} \notag \\
\		\doteq \rho^{-K(1-2r)}  \notag
\end{align}
where $\epsilon_{2}=\frac{\log(K\lambda_{{\rm max}})}{\log\rho}$. We have thus shown that
\begin{align}
	\rho^{-K(1-2r)} \ \dotleq \ \prob{\mathcal{J}} \ \dotleq \ \rho^{-K(1-2r)}  \notag
\end{align}
and hence
\begin{align}
\prob{\mathcal{J}} = P_{J} (\rho,r) \ \doteq \ \rho^{-K(1-2r)} . \notag
\end{align}
Since $P_{J}(\rho,r) \leq P_{\mathcal{O}}(\rho,r)$ as a result of (\ref{jensenmutualinfo2}), and since the outage probability is a lower bound to the error probability achieved by any code \cite{zheng_tradeoff}, we have 
\begin{align}
\label{finishingt}
P_{J}(\rho,r) \leq P_{\mathcal{O}}(\rho,r) \leq P_{e}(\rho,r) {\rm .}
\end{align}
Following the approach introduced in \cite{coronel06}, we now complete the proof of the theorem by identifying  a family of codes which has
$P_{e}(\rho,r)\  \doteq \ P_{J}(\rho,r)$ and hence results in a DM-tradeoff curve which equals the ``Jensen'' DM-tradeoff curve derived above.
We start by writing
\begin{align}
P_{e}(\rho,r) =&  \ \prob{\mathcal{J}}\prob{{\rm error} | \mathcal{J}} + \prob{{\rm error}, \bar{\mathcal{J}}} \notag \\ 
			\leq& \ \prob{\mathcal{J}}+ \prob{{\rm error}, \bar{\mathcal{J}}}. \notag
\end{align}
Next, we upper-bound $\prob{{\rm error}, \bar{\mathcal{J}}}$ through the union bound
 \begin{align} 
 \label{unionbound}
\prob{{\rm error}, \bar{\mathcal{J}}} \leq \rho^{2Nr} \ \prob{\hat{\mathbf{x}}\rightarrow\tilde{\mathbf{x}}, \bar{\mathcal{J}}} \end{align}
where we used the fact that the codebook, $\mathcal{C}_{r}(\rho)$, contains $\rho^{2Nr}$ codewords and $\prob{\hat{\mathbf{x}}\rightarrow\tilde{\mathbf{x}}, \bar{\mathcal{J}}}$ denotes the maximum pairwise error probability (over all codeword pairs and all channels in $\bar{\mathcal{J}}$) for ML decoding. With $\Delta\mathbf{x}=\tilde{\mathbf{x}}-\hat{\mathbf{x}}$, we have
\begin{align}
& \prob{\hat{\mathbf{x}}\rightarrow \tilde{\mathbf{x}} | \heff  = \mathbf{H}} =  Q\left(\sqrt{\frac{\rho}{2}}{||\mathbf{H}\Delta \mathbf{x} ||_{F}}\right) \notag \\
&			\ 		 \leq  \exp\left[-\frac{\rho}{4}{||\mathbf{H}\Delta \mathbf{x}||^2_{F}}\right]  \ \dotleq  \ \exp\!\left[-\frac{\rho}{4(1+K)}\|\mathbf{\Phi}(\Delta\mathbf{x})\tilde{\mathbf{h}}\|^2_{F}\right] \notag \\
&			\ \label{phistep2}				\leq \exp\left[-\frac{\rho}{4(1+K)} \mu_{{\rm min}}(\rho)\|\tilde{\mathbf{h}}\|^2 \right] 
\end{align} 
where $\mu_{{\rm min}}(\rho)$ denotes the minimum eigenvalue of $\left(\mathbf{\Phi}(\Delta\mathbf{x})\right)^{H}\mathbf{\Phi}(\Delta\mathbf{x})$ and (\ref{phistep2}) follows from applying the Rayleigh-Ritz theorem.  Substituting $\|\tilde{\mathbf{h}}\|^2 =   \sum_{i=1}^{K}\rho^{-v_{i}-u_{i}}$ into (\ref{phistep2}), we have
\begin{align} 
 \prob{\hat{\mathbf{x}}\rightarrow \tilde{\mathbf{x}}, \bar{\mathcal{J}}}  \notag 
 &\leq \mathbb{E}_{\tilde{\mathbf{h}}\in\bar{\mathcal{J}} }\left\{  \exp\left[-\frac{\mu_{{\rm min}}(\rho)}{4(1+K)} \sum_{i=1}^{K}\rho^{1-v_{i}-u_{i}} \right]  \right\}  \notag \\
&  \label{PXXXX1}\leq    \exp\left[-\frac{\mu_{{\rm min}}(\rho)}{4(1+K)}\rho^{2r} \right]   
\end{align}
where  (\ref{PXXXX1}) follows since the event $\bar{\mathcal{J}}$ requires that $\sum_{i=1}^{K} \rho^{1-v_{i}-u_{i}} \geq \rho^{2r}$.  Finally, inserting (\ref{PXXXX1}) into 
(\ref{unionbound}),  we get 
\begin{align}
\label{error-no-jensen} \prob{{\rm error}, \bar{\mathcal{J}}}\  \dotleq \ \rho^{2Nr} \exp\left[-\frac{\mu_{{\rm min}}(\rho)}{4(1+K)}\rho^{2r} \right]  {\rm .}
\end{align}
The proof is complete since $\mathbf{\Phi}(\Delta\mathbf{x})$ has full rank for all $\Delta\mathbf{x}$ and for all codebooks in $\mathcal{C}_{r}$ and hence $\mu_{{\rm min}}(\rho)>0$ which implies that (\ref{error-no-jensen}) decays exponentially in $\rho$ for all $r \in [0,1/2]$. Summarizing our results, we obtain
\begin{align}
P_{e}(\rho,r) 	\leq&  \ \prob{\mathcal{J}} + \prob{{\rm error}, \bar{\mathcal{J}}}  \notag \\
                      \dotleq & \ \prob{\mathcal{J}} + \rho^{2Nr} \exp\left[-\frac{\mu_{{\rm min}}(\rho)}{4(1+K)}\rho^{2r} \right]   \notag \\
                      \dotleq & \ \prob{\mathcal{J}} = P_{J} (\rho,r)  \notag
\end{align}
which combined with (\ref{finishingt}) yields the desired result. 

\bibliography{IEEEabrv,confs-jrnls,publishers,cebib}

\begin{thebibliography}{10}
\providecommand{\url}[1]{#1}
\csname url@rmstyle\endcsname
\providecommand{\newblock}{\relax}
\providecommand{\bibinfo}[2]{#2}
\providecommand\BIBentrySTDinterwordspacing{\spaceskip=0pt\relax}
\providecommand\BIBentryALTinterwordstretchfactor{4}
\providecommand\BIBentryALTinterwordspacing{\spaceskip=\fontdimen2\font plus
\BIBentryALTinterwordstretchfactor\fontdimen3\font minus
  \fontdimen4\font\relax}
\providecommand\BIBforeignlanguage[2]{{%
\expandafter\ifx\csname l@#1\endcsname\relax
\typeout{** WARNING: IEEEtran.bst: No hyphenation pattern has been}%
\typeout{** loaded for the language `#1'. Using the pattern for}%
\typeout{** the default language instead.}%
\else
\language=\csname l@#1\endcsname
\fi
#2}}

\bibitem{laneman_st}
J.~N. Laneman and G.~W. Wornell, ``Distributed space-time-coded protocols for
  exploiting cooperative diversity in wireless networks,'' \emph{IEEE Trans.
  Inf. Theory}, vol.~49, no.~10, pp. 2415 -- 2425, Oct. 2003.

\bibitem{jing06}
Y.~Jing and B.~Hassibi, ``Distributed space-time coding in wireless relay
  networks,'' \emph{IEEE Trans. Wireless Comm.}, vol.~5, no.~12, pp.
  3524--3536, Dec. 2006.

\bibitem{azarian05}
K.~Azarian, H.~{El~Gamal}, and P.~Schniter, ``On the achievable
  diversity-multiplexing tradeoff in half-duplex cooperative channels,''
  \emph{IEEE Trans. Inf. Theory}, vol.~51, no.~12, pp. 4152--4172, Dec. 2005.

\bibitem{nabarj}
R.~U. Nabar, H.~B\"olcskei, and F.~W. Kneub\"uhler, ``Fading relay channels:
  {P}erformance limits and space-time signal design,'' \emph{IEEE J. Sel. Areas
  Comm.}, vol.~22, no.~6, pp. 1099--1109, Aug. 2004.

\bibitem{hammerstroem04}
I.~Hammerstr\"om, M.~Kuhn, and A.~Wittneben, ``Cooperative diversity by relay
  phase rotations in block fading environments,'' in \emph{Proc. Fifth IEEE
  Workshop on Signal Processing Advances in Wireless Communications (SPAWC)},
  July 2004, pp. 293--297.

\bibitem{kuppinger06}
P.~Kuppinger, ``Transformation of distributed spatial into temporal diversity
  by relay phase rotations,'' M.Sc. Thesis, Imperial College London, Sept.
  2006.

\bibitem{slimane06}
S.~B. Slimane and A.~Osseiran, ``Relay communication with delay diversity for
  future communication systems,'' in \emph{Proc. IEEE VTC (Fall)}, Sept. 2006,
  pp. 1--5.

\bibitem{zheng_tradeoff}
L.~Zheng and D.~N.~C. Tse, ``Diversity and multiplexing: {A} fundamental
  tradeoff in multiple-antenna channels,'' \emph{IEEE Trans. Inf. Theory},
  vol.~49, no.~5, pp. 1073--1096, May 2003.

\bibitem{coronel06}
P.~Coronel and H.~B{\"o}lcskei, ``Diversity-multiplexing tradeoff in
  selective-fading {MIMO} channels,'' in \emph{Proc. IEEE ISIT}, Nice, France,
  June 2007, to appear.

\bibitem{tavildar06}
S.~Tavildar and P.~Viswanath, ``Approximately universal codes over slow-fading
  channels,'' \emph{IEEE Trans. Inf. Theory}, vol.~52, no.~7, pp. 3233--3258,
  July 2007.

\bibitem{gastpar05}
M.~Gastpar and M.~Vetterli, ``On the capacity of large {Gaussian} relay
  networks,'' \emph{IEEE Trans. Inf. Theory}, vol.~51, no.~3, pp. 765--779,
  March 2005.

\bibitem{Bol06}
H.~B\"olcskei, R.~U. Nabar, {\"O}.~Oyman, and A.~J. Paulraj, ``Capacity scaling
  laws in {MIMO} relay networks,'' \emph{IEEE Trans. Wireless Comm.}, vol.~5,
  no.~6, pp. 1433--1444, Jun. 2006.

\bibitem{Horn85}
R.~A. Horn and C.~R. Johnson, \emph{Matrix Analysis}.\hskip 1em plus 0.5em
  minus 0.4em\relax New York, NY: Cambridge Press, 1985.

\bibitem{Salo06}
J.~Salo, H.~El-Sallabi, and P.~Vainikainen, ``The distribution of the product
  of independent {Rayleigh} random variables,'' \emph{IEEE Trans. Ant. and
  Prop.}, vol.~54, no.~2, pp. 639--643, Feb. 2006.

\bibitem{Abr65}
M.~Abramowitz and I.~Stegun, \emph{Handbook of Mathematical Functions}.\hskip
  1em plus 0.5em minus 0.4em\relax New York: Dover, 1965.

\end{thebibliography}
\bibliographystyle{IEEEtran}

\end{document}